\documentclass[12pt]{iopart}
\usepackage{iopams} 
\usepackage{graphics}
\usepackage{pennames}
\usepackage{amssymb}
\usepackage{setstack}
\def\La{\Lambda}

\def\XI{\Xi^{-}}

\newcommand{\decayarrow}{\makebox[0mm][l]{\rule{0.33em}{0mm}\rule[0.55ex]{0.044em}{1.55ex}}\rightarrow}
\begin{document}
\title[Rapidity distributions of strange particles in Pb-Pb at 158 $A$\ GeV/$c$]
{Rapidity distributions around mid-rapidity of strange particles in Pb-Pb collisions at 158 $A$ GeV/$c$}
\author{  
F~Antinori$^{k}$,
P~Bacon$^{e}$,
A~Badal{\`a}$^{f}$,
R~Barbera$^{f}$,
A~Belogianni$^{a}$,
I~J~Bloodworth$^{e}$,
M~Bombara$^{h}$,
G~E~Bruno$^{b}$
\footnote[1]{To
whom correspondence should be addressed (giuseppe.bruno@ba.infn.it)}
,
S~A~Bull$^{e}$,
R~Caliandro$^{b}$,
M~Campbell$^{g}$,
W~Carena$^{g}$,
N~Carrer$^{g}$,
R~F~Clarke$^{e}$,
A~Dainese$^{k}$,
D~Di~Bari$^{b}$,
S~Di~Liberto$^{n}$,
R~Divi\`a$^{g}$,
D~Elia$^{b}$,
D~Evans$^{e}$,
G~A~Feofilov$^{p}$,
R~A~Fini$^{b}$,
P~Ganoti$^{a}$,
B~Ghidini$^{b}$,
G~Grella$^{o}$,
H~Helstrup$^{d}$,
K~F~Hetland$^{d}$,
A~K~Holme$^{j}$,
A~Jacholkowski$^{f}$,
G~T~Jones$^{e}$,
P~Jovanovic$^{e}$,
A~Jusko$^{e}$,
R~Kamermans$^{r}$,
J~B~Kinson$^{e}$,
K~Knudson$^{g}$,
V~Kondratiev$^{p}$,
I~Kr\'alik$^{h}$,
A~Krav\v c\'akov\'a$^{i}$,
P~Kuijer$^{r}$,
V~Lenti$^{b}$,
R~Lietava$^{e}$,
G~L\o vh\o iden$^{j}$,
V~Manzari$^{b}$,
M~A~Mazzoni$^{n}$,
F~Meddi$^{n}$,
A~Michalon$^{q}$,
M~Morando$^{k}$,
P~I~Norman$^{e}$,
A~Palmeri$^{f}$,
G~S~Pappalardo$^{f}$,
B~Pastir\v c\'ak$^{h}$,
R~J~Platt$^{e}$,
E~Quercigh$^{k}$,
F~Riggi$^{f}$,
D~R\"ohrich$^{c}$,
G~Romano$^{o}$,
K~\v{S}afa\v{r}\'{\i}k$^{g}$,
L~\v S\'andor$^{h}$,
E~Schillings$^{r}$,
G~Segato$^{k}$,
M~Sen\'e$^{l}$,
R~Sen\'e$^{l}$,
W~Snoeys$^{g}$,
F~Soramel$^{k}$
\footnote[2]{Permanent
address: University of Udine, Udine, Italy},
M~Spyropoulou-Stassinaki$^{a}$,
P~Staroba$^{m}$,
R~Turrisi$^{k}$,
T~S~Tveter$^{j}$,
J~Urb\'{a}n$^{i}$,
P~van~de~Ven$^{r}$,
P~Vande~Vyvre$^{g}$,
A~Vascotto$^{g}$,
T~Vik$^{j}$,
O~Villalobos~Baillie$^{e}$,
L~Vinogradov$^{p}$,
T~Virgili$^{o}$,
M~F~Votruba$^{e}$,
J~Vrl\'{a}kov\'{a}$^{i}$\ and
P~Z\'{a}vada$^{m}$
}
\address{
$^{a}$ Physics Department, University of Athens, Athens, Greece\\
$^{b}$ Dipartimento IA di Fisica dell'Universit{\`a}
       e del Politecnico di Bari and INFN, Bari, Italy \\
$^{c}$ Fysisk Institutt, Universitetet i Bergen, Bergen, Norway\\
$^{d}$ H{\o}gskolen i Bergen, Bergen, Norway\\
$^{e}$ University of Birmingham, Birmingham, UK\\
$^{f}$ University of Catania and INFN, Catania, Italy\\
$^{g}$ CERN, European Laboratory for Particle Physics, Geneva, Switzerland\\
$^{h}$ Institute of Experimental Physics, Slovak Academy of Science,
              Ko\v{s}ice, Slovakia\\
$^{i}$ P.J. \v{S}af\'{a}rik University, Ko\v{s}ice, Slovakia\\
$^{j}$ Fysisk Institutt, Universitetet i Oslo, Oslo, Norway\\
$^{k}$ University of Padua and INFN, Padua, Italy\\
$^{l}$ Coll\`ege de France, Paris, France\\
$^{m}$ Institute of Physics, Prague, Czech Republic\\
$^{n}$ University ``La Sapienza'' and INFN, Rome, Italy\\
$^{o}$ Dipartimento di Scienze Fisiche ``E.R. Caianiello''
       dell'Universit{\`a} and INFN, Salerno, Italy\\
$^{p}$ State University of St. Petersburg, St. Petersburg, Russia\\
$^{q}$ IReS/ULP, Strasbourg, France\\
$^{r}$ Utrecht University and NIKHEF, Utrecht, The Netherlands
}
\begin{abstract}
The production at central rapidity of \PKzS, \PgL, $\Xi$\ and  $\Omega$\ particles  
in Pb-Pb collisions at 158 $A$\ GeV/$c$\ has been measured   
by the NA57 experiment over a centrality range corresponding to   
%53\% of 
the most central 53\% of the inelastic Pb-Pb cross section.   
In this paper we present the rapidity distribution of each particle   
%for high purity samples of these particles  
in the central rapidity unit  
as a function of the event centrality.  
%Hydrodynamical models are fitted to  
%the rapidity distributions %are fitted to hydrodynamical models 
%to study the longitudinal dynamics of the Pb-Pb collisions.  
The distributions are analyzed based on hydrodynamical models of the 
collisions. 
%\PKzS\ yields in one unit of rapidity are presented as a function of centrality.   
%for a sample of events corresponding to the most central  53\% of the inelastic cross section.   
\end{abstract}
\vspace{-0.6cm}
\submitto{\JPG}
%Version.4  12/07/2005
%\maketitle
%
\section{Introduction} 
Lattice quantum chromodynamic calculations predict a new state of matter of deconfined 
quark and gluons (quark gluon plasma, QGP) at an energy density exceeding $\sim$\ 1 
GeV/fm$^3$~\cite{lattice}.  
Nuclear matter at high energy density has been extensively studied through %high energy  
ultra-relativistic  
heavy ion collisions (for recent developements, see reference~\cite{QM04}). 
%In collisions of heavy nuclei at ultra-relativistic energies nuclear matter is  
%%compressed and significantly excited.
%strongly compressed and heated. 
%%The estimated energy density reached is about 3 GeV/fm$^3$.
%%
%%reaching energy density of about 3 GeV/fm$^3$.  
%In such extreme conditions novel physical  
%phenomena can arise, and it is 
%%believed that the excited nuclear matter
%expected that such an excited nuclear matter 
%undergoes  
%a phase transition into a system of deconfined quarks and gluons  
%(quark-gluon plasma, QGP)~\cite{QGP}. 

Within the experimental programme with heavy-ion beams at CERN SPS,
NA57 is a dedicated experiment for the study of  
the production of strange and multi-strange particles
in Pb-Pb %and p-Be 
collisions at mid-rapidity~\cite{NA57proposal}. 
%It has continued the WA97 measurements by extending them over a wider centrality 
%range and to lower (40 $A$\ GeV/$c$) beam energy.

The measurement of strange particle production provides one of the most powerful tools    
to study the dynamics of the reaction. %in heavy-ion collisions. 
In particular, an enhanced production of strange particles in nucleus--nucleus collisions with respect  
to proton--induced reactions was suggested long ago as a possible signature of the 
phase transition from colour confined hadronic matter to a QGP~\cite{StrEnh}. 
The enhancement %has to  
%was predicted to
%was found to 
is expected to  
increase with the strangeness content of the hyperon. 
These %predictions 
features 
were first observed by the WA97 experiment~\cite{WA97Enh} and subsequently confirmed and studied in more detail  
by the NA57 experiment~\cite{NA57Enh}.   
%\newline
Other insights into the reaction dynamics have been inferred  
from the $p_{\tt T}$\ distributions of 
the strange particles: the study of the transverse expansion of the collision and the 
$p_{\tt T}$\ dependence of the nuclear modification factors have been presented, respectively,  
in reference~\cite{BlastPaper} and~\cite{RcpPaper}.   
%Other insights into the reaction dynamics have been inferred  
%from the $p_{\tt T}$\ distributions of   
%the strange particles~\cite{BlastPaper,RcpPaper}.  
%both for the initial stage of the reaction and for its late developement 
%by looking respectively to the soft ()
%
%\newline

Rapidity distributions provide a tool to study the longitudinal dynamics;  
%in particular, 
e.g.,  
differences between protons and anti-protons have been interpreted as 
a consequence of the nuclear stopping~\cite{Busza}.  
%If the hyperons keep  `memory' of the initial baryon density as protons do,  then  
%a similar pattern as for the net-protons would emerge also for the `net-hyperon'.   
%
%If hyperons keep a `memory' of the initial baryon density, like protons, a similar pattern  
%for the rapidity distribution of net-protons and `net-hyperons' should emerge.  
If hyperons, like protons, keep a `memory' of the initial baryon density, then the {\em relative} 
pattern for the rapidity distribution of hyperons and anti-hyperons should resemble    
that of protons and anti-protons~\cite{NA49netprot}.  

Hydrodynamical properties of the expanding matter created in heavy ion reactions have been 
discussed by Landau~\cite{Landau} and Bjorken~\cite{Bjorken} in theoretical pictures using 
different initial conditions.  In both scenarios, thermal equilibrium is quickly achieved and 
the subsequent isentropic expansion is governed by hydrodynamics.   
%
%\newline

%The energy dependence of the K/$\pi$\ ratios has been suggested as a possible 
%signal of the onset of the chiral transition (for the latest, see reference~\cite{Horn}). 
%In this respect, precise measurements of $dN/dy$\ for \PKzS\ at low (40) and 
%top (160 A GeV/$c$) SPS energies may %reveal 
%result particularly  
%useful.  
%\newline

%Last but not least, the relative particle abundances, which can be 
%%computed 
%deduced  
%from the 
%rapidity densities %$\frac{dN}{dy}  \left|_{y=y_{cm}}  \right.$, 
%$\frac{dN}{dy} $,  
%are the basic ingredient 
%of the thermal models~\cite{ThermalModel1,ThermalModel2,ThermalModel3,ThermalModel4}.  
%Precise measurements of the strange particle 
%%$\frac{dN}{dy}  \left|_{y=y_{cm}} \right.$\ 
%$\frac{dN}{dy} $\  
%%would put stronger 
%put strong 
%constrains on the thermal fits. This in turn  
%%would also help 
%could help 
%in the search 
%for signs of non-equilibrium 
%phenomena 
%(see for instance reference~\cite{Antinori}), 
%possibly seen in small deviations from the thermal model expectations.  
%%where one has to look at fine details.
\section{Data sample}
The NA57 experiment has been described in detail elsewhere~\cite{BlastPaper,NA57descr}.  
The results presented in this paper are based on the analysis of 
the Pb-Pb data sample collected %in Pb-Pb collisions, 
at 158 $A$\ GeV/$c$ %,  
%consisting of 460 M events. %at 158 $A$\ GeV/$c$.  
(460 M events); 
the sample analyzed here 
corresponds to 
%about 
the most central  %60\% 
53\%  
of the inelastic Pb-Pb cross-section.  
%The triggered fraction of the Pb-Pb inelastic cross-section is about 60\%. 
%
%All data are corrected for geometrical acceptance and for detector and 
%reconstruction inefficiencies on an event-by-event  
%basis, with the procedure described in reference~\cite{QM02Manzari}. 
%
The data have been divided into five centrality classes, labelled as 0,1,2,3 and 4 
from the most peripheral to the most central, according to the value of the charged particle  
multiplicity sampled at central rapidity by a %Silicon Microstrip Multiplicity Detector. 
dedicated silicon microstrip detector.   

The procedure to measure the multiplicity distribution and   
to determine the centrality of the collision for each class  
is described in reference~\cite{Multiplicity}. 
The fractions of the inelastic cross-section for each of the five classes, 
%calculated assuming an inelastic Pb-Pb cross-section of 7.26 barn, 
determined assuming an inelastic Pb-Pb cross-section of 7.26 barn
(as calculated with the Glauber model),  
are given in table~\ref{tab:centrality}.  
\begin{table}[h]
\caption{Centrality ranges for the five classes.
\label{tab:centrality}}
\begin{center}
\begin{tabular}{llllll}
\hline
 Class &   $4$   &   $3$   &   $2$   &  $1$   &   $0$ \\ \hline
 $\sigma/\sigma_{inel}$\ \; (\%)  & 0 to 4.5 & 4.5 to 11  & 11 to 23 & 23 to 40  & 40 to 53  \\
 \hline
\end{tabular}
\end{center}
\end{table}
\noindent

\section{Analysis and results}
A detailed description of the particle selection procedure, as well as 
%the applied corrections 
of the corrections  
for geometrical acceptance and for detector and reconstruction 
inefficiencies, can be 
found in reference~\cite{BlastPaper}. 
Here we describe the principal features of the analysis.  
%\newline 

The selection of strange particles was accomplished by reconstructing their weak decays 
into final states containing only charged particles: 
\begin{equation}
\label{eq:decay}
\begin{array}{lllllll}
    \PKzS &\rightarrow & \pi^+ + \pi^-  & \hspace{10mm}&
    \La  &\rightarrow & p + \pi^-                             \\ \\
    \XI &\rightarrow & \La + \pi^- &  \hspace{10mm}&  \Omega^- &\rightarrow &
   \La + K^-  \\
            &            &
   \decayarrow  p + \pi^-  &  &  &  & \decayarrow  p + \pi^-   \\
\end{array}
\end{equation}
and their corresponding charge conjugates for antihyperons. 
Geometric and kinematic constraints were used 
%to isolate the strange particle species in high purity samples: 
to obtain high purity strange particle samples:  
the total amount of combinatorial background is estimated to be 
0.7\%, 0.3\% and 1.2\% for \PKzS, \PgL\ and \PagL\ respectively, and less than 5\% 
%and 6\% 
for $\Xi$\ and $\Omega$.  
%, respectively.   
%\newline

An event by event procedure, using a GEANT-based~\cite{GEANT} Monte Carlo simulation of 
the apparatus, was applied to correct the data for geometrical 
acceptance and for detector and reconstruction inefficencies. This  
procedure, which is very accurate but rather time consuming, assigns a weight to each 
reconstructed particle and was specifically  
developed for rare signals such as $\Omega$\ and $\Xi$. The same procedure was also applied 
to small fractions of the much more abundant \PKzS\ and $\Lambda$\ samples.     
The selected particles have been sampled uniformly over the whole data taking period;    
the sizes of those  
subsamples were chosen in order to reach a statistical accuracy better than the limits imposed 
by the systematic errors.   
%\newline

\begin{figure}[hbt]
\centering
\resizebox{0.70\textwidth}{!}{%
\includegraphics{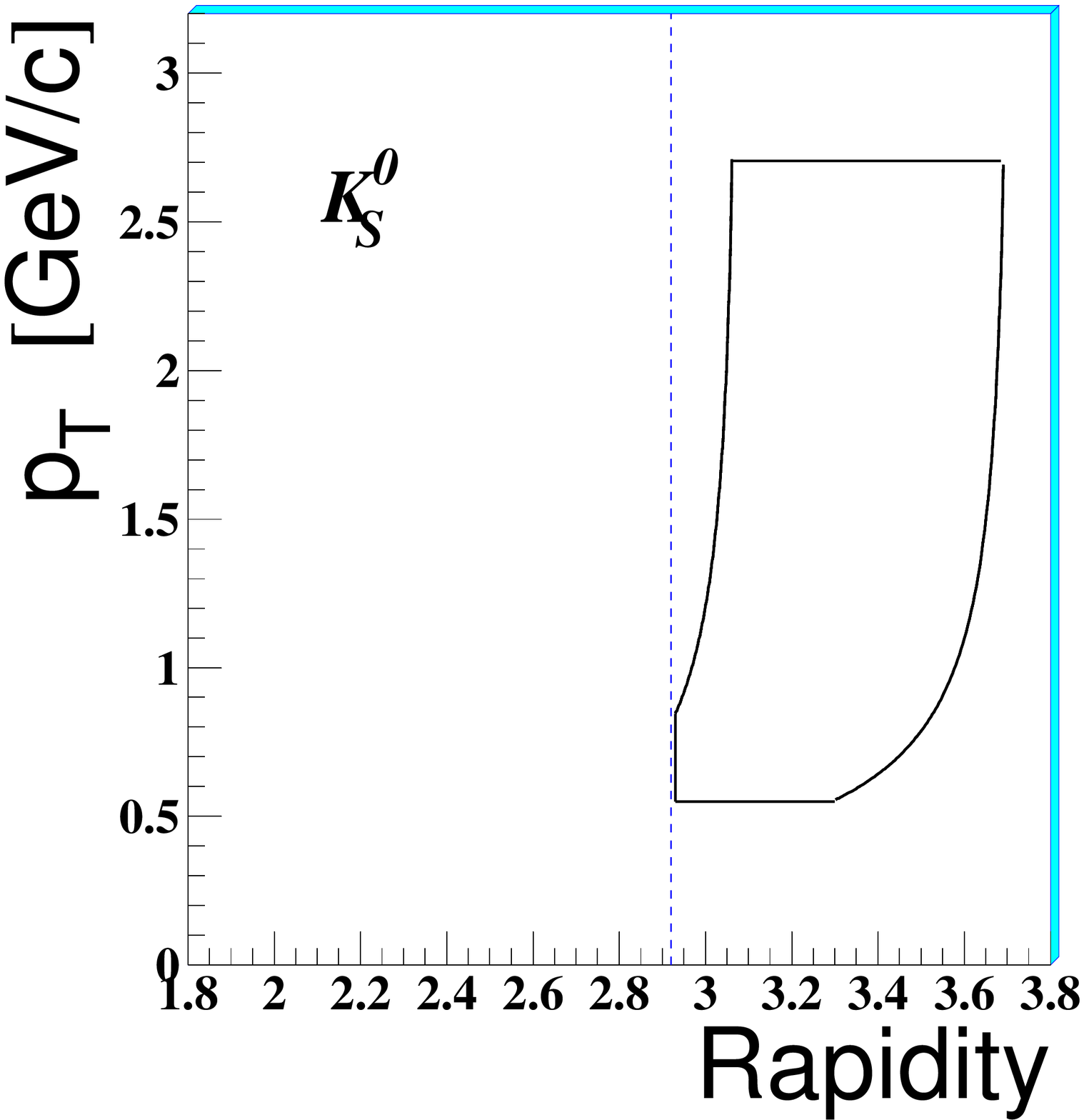}
\includegraphics{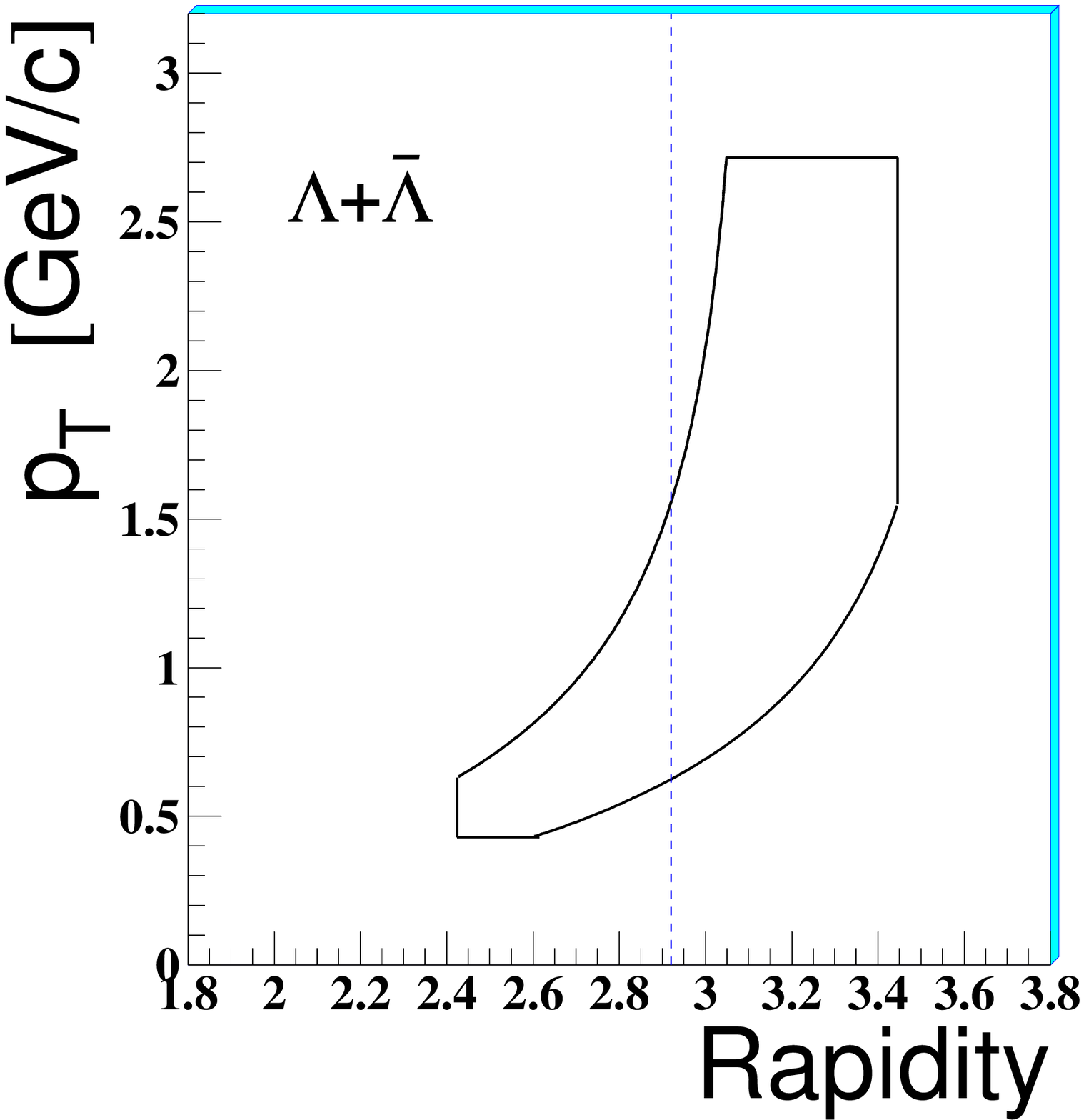}}\\
\resizebox{0.70\textwidth}{!}{%
\includegraphics{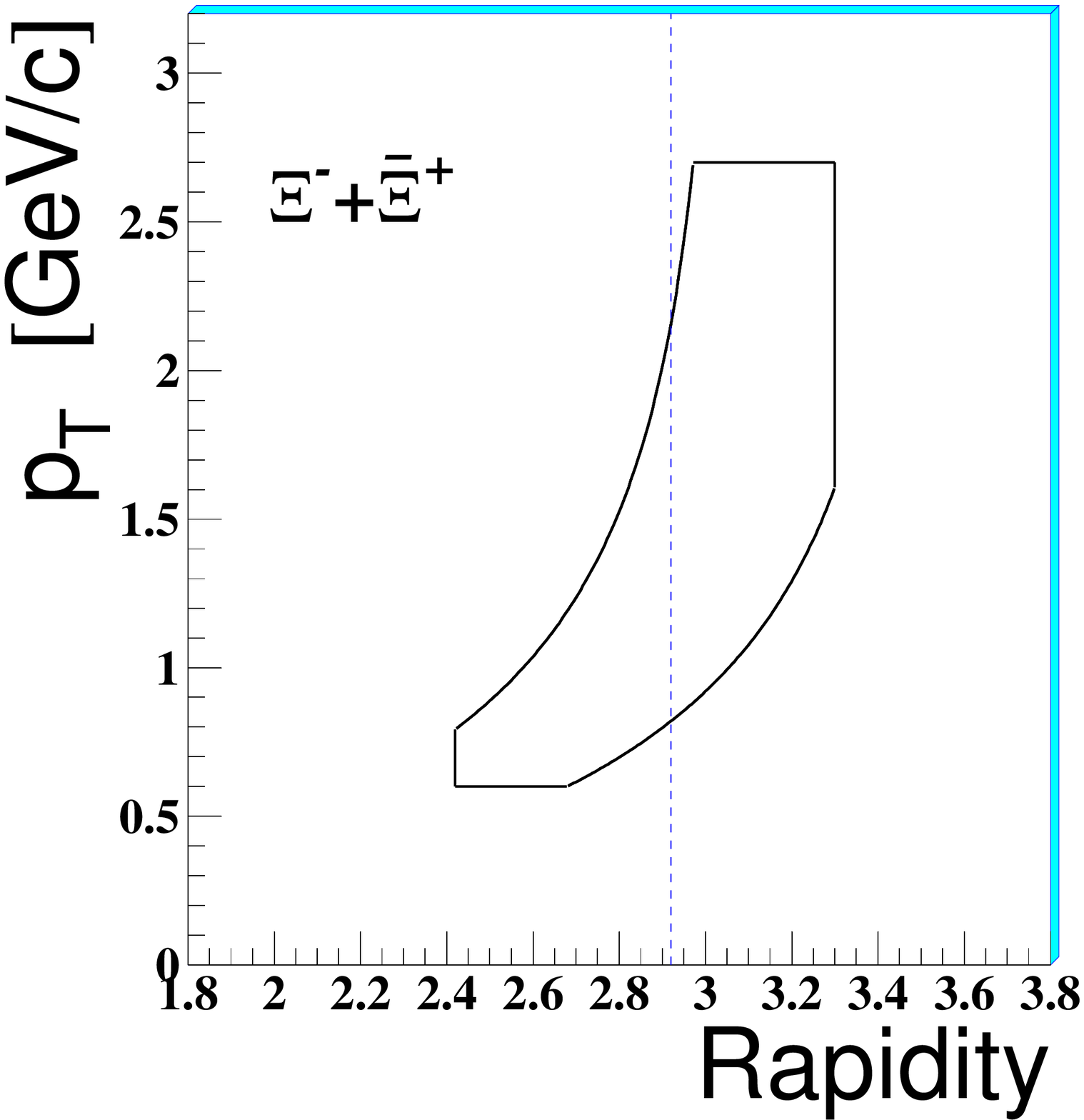}
\includegraphics{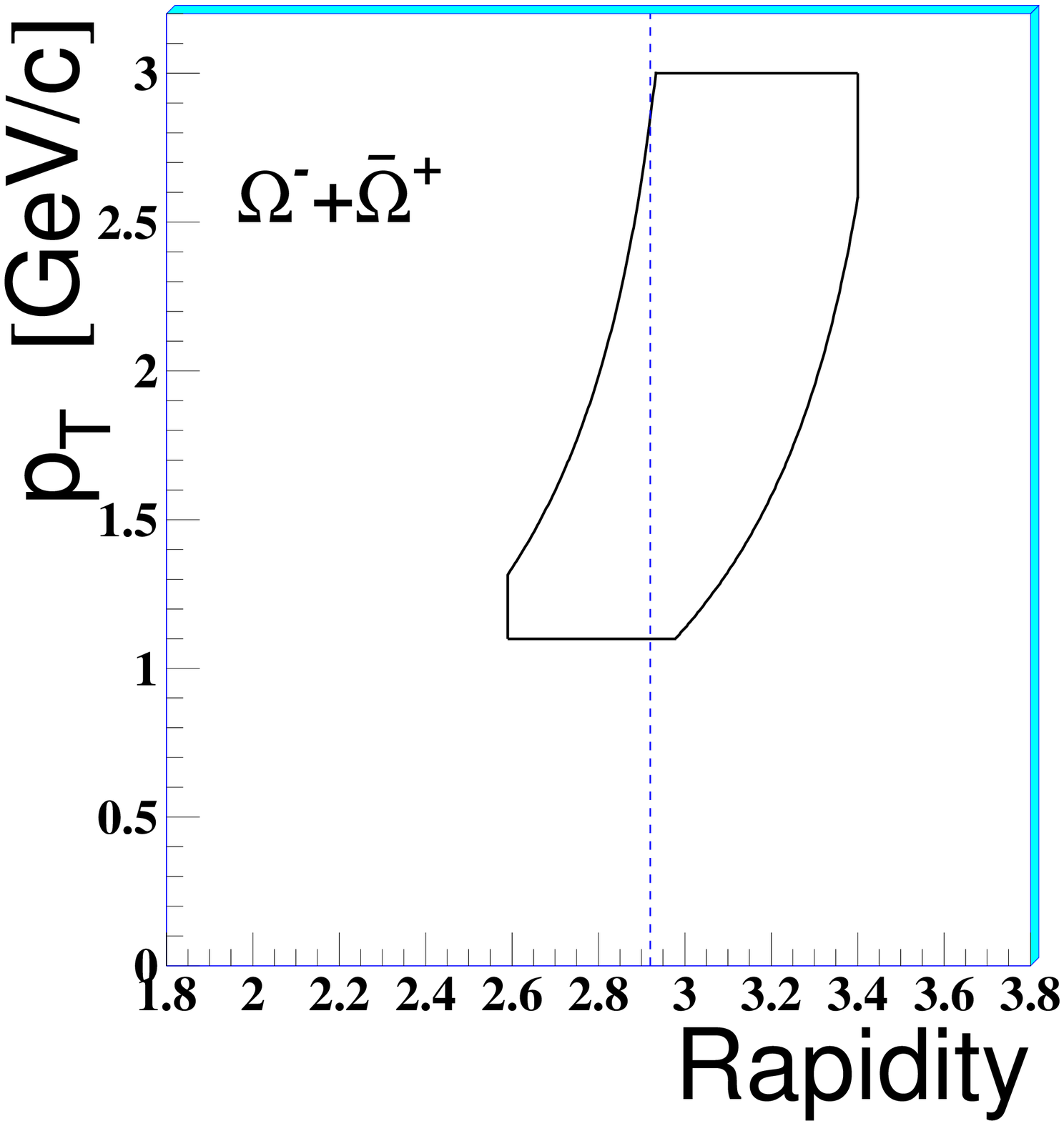}}
\caption{\rm The $y$--$p_{\tt T}$\
         acceptance windows. %superimposed to the data samples.
         Dashed lines show the position of mid-rapidity ($y_{cm}=2.92$).}
\label{fig:acceptance}
\end{figure}

%For each particle species a fiducial acceptance window was defined using the Monte Carlo  
%simulation of the apparatus, in order to exclude the borders where the corrections can  
%be evaluated with less accuracy. 
Acceptance regions have been refined off-line in such a way 
to  excluded from the final sample the border regions where the correction 
factors would have larger systematic errors.     
The selected regions of the transverse momentum vs. rapidity plane 
are shown in figure~\ref{fig:acceptance}.  

Extensive checks were performed % on on  the simulation 
by comparing real and Monte Carlo distributions for %several parameters, 
several geometrical and kinematical parameters, 
%in particular those used for particle sample selection, 
in particular those used in the particle selection procedure. 
%and they  show good agreement~\cite{BlastPaper,Fanebust}.
In all cases a good agreement between MC and %the 
data has been found~\cite{BlastPaper,Fanebust}.  

The double-differential %$(y,p_{\tt T})$\ 
distribution for each particle species has been parameterized using the expression  
\begin{equation} 
\label{eq:expo} 
\frac{d^2N}{p_{\tt T}dp_{\tt T} dy} %=
                  \propto
                  \frac{dN}{dy} \hspace{1mm} 
                       \cdot \exp\left(-\frac{\sqrt{m^2+p_{\tt T}^2}}{T_{app}}\right)
\end{equation}
where the inverse slope parameter $T_{app}$\   
(`apparent temperature') has been extracted by means of a maximum likelihood fit of  
equation~\ref{eq:expo} to the data~\cite{BlastPaper}.  
By using equation~\ref{eq:expo} we can extrapolate the rapidity distribution measured 
in the selected acceptance window down to $p_{\tt T}=0$:
\begin{equation}
\frac{dN}{dy}=\int_{0}^{\infty} {\rm d}p_{\tt T}  
  \frac{{\rm d}^2N}{{\rm d}p_{\tt T} {\rm d}y}.
\label{eq:yield}
\end{equation}
\noindent
The contribution of the extrapolation procedure to the systematic errors 
on $\frac{dN}{dy}$\ has been evaluated using an alternative 
parameterization for the $p_{\tt T}$\ dependence of the invariant 
double-differential distribution: instead of the single exponential 
in equation~\ref{eq:expo}, the blast-wave model, a %hydrodynamically motivated 
hydrodynamics-inspired  
model with a kinetic freeze-out temperature $T_f$\ and a linear transverse flow 
velocity field $\beta_{\perp}(r)$~\cite{BlastModel},  
was used  with freeze-out   
parameters %taken from reference~\cite{BlastPaper}.   
$T_f=144$\ MeV and $<\beta_{\perp}>=0.38$\    
which we determined from the blast-wave analysis of the $p_{\tt T}$\ 
spectra~\cite{BlastPaper}.   
This alternative procedure does not affect the shape of the $\frac{dN}{dy}$\ 
distributions but only the absolute values of the extracted yields; 
the yield of the $\PKzS$\ meson increases by $ 6\%$, that of the \PgL\ hyperon
remains unchanged, 
those of the $\Xi$\ and $\Omega$\ hyperons decrease by  
$5\%$\ and $20\%$, respectively. This results in the   
main source of systematic uncertainty for the $\Omega$\ hyperon.  
\subsection{Results in the centrality range 0-53\%}
\begin{figure}[hbt]
\centering
\resizebox{1.00\textwidth}{!}{%
\includegraphics{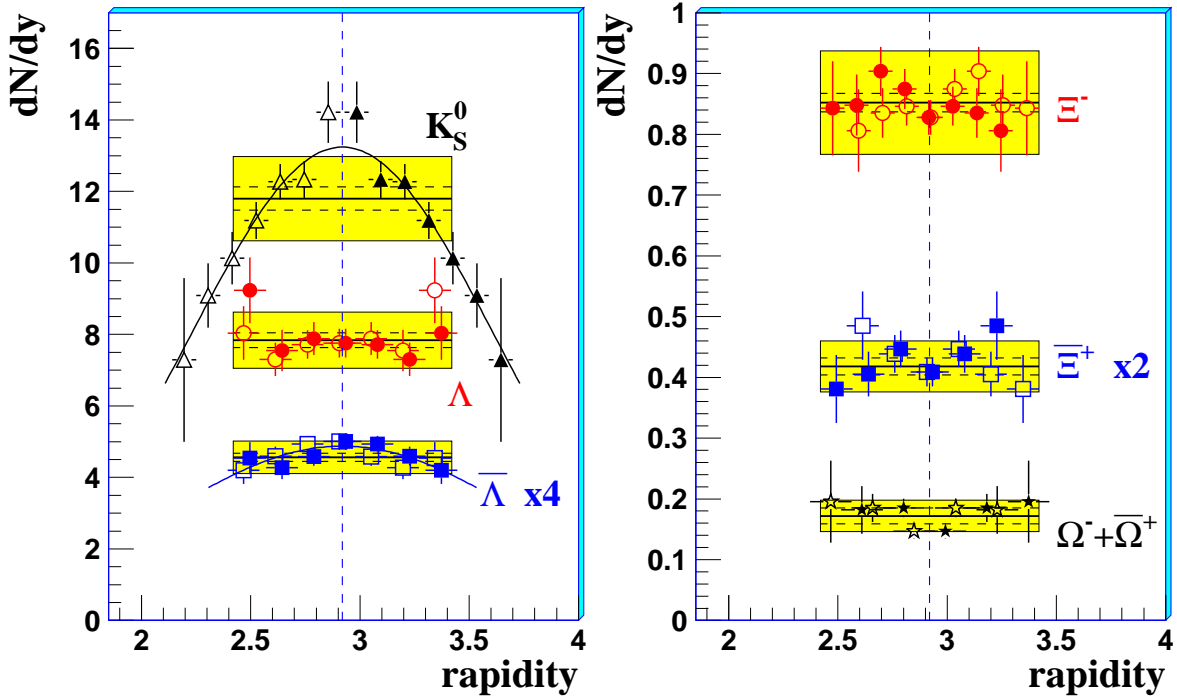}
}
\caption{Rapidity distributions of strange particles in the most central 53\% of  
        Pb-Pb interactions at 158 $A$\ GeV/$c$. Closed symbols are measured data, open 
	symbols are measured points reflected around mid-rapidity. 
        The \PagL\ and \PagXp\ results have been scaled by factors 4 and 2, respectively,  
        for %better displaying.   
             dispaly purposes.  
        The superimposed boxes show the yields measured in one unit of rapidity 
	(as  published in references~\cite{NA57Enh,EneDepPaper}) with the dashed and full 
	lines indicating the statistical and systematic errors, respectively.}  
\label{fig:yspectra_160}
\end{figure}

The measured rapidity distributions for the centrality range corresponding 
to the most central 53\% of the Pb-Pb inelastic cross-section (total sample)
are shown in figure~\ref{fig:yspectra_160} with closed symbols.  
For all  hyperons the rapidity distributions are found to be symmetric
with respect to %mid-rapidity 
the rapidity of the centre of mass (`mid-rapidity') within the statistical errors.  
A similar conclusion cannot be drawn for \PKzS\ since our acceptance
coverage does not extend to  backward rapidity.
The symmetry of the Pb-Pb colliding system allows us to reflect 
the rapidity distributions around mid-rapidity.   
In figure~\ref{fig:yspectra_160} open symbols are used to plot the measured points after 
such a reflection.  
%reflection around mid-rapidity.  
%.   
%This finding gives us confidence in the $(p_{\tt T},y)$\ factorization of the double differential  
%cross-section which has been assumed in equation~\ref{eq:expo}. 
%\newline

The rapidity distributions of \PgL, \PgXm, \PagXp and $\Omega$\ %are found to be flat 
are compatible, within the error bars, with being flat 
within the NA57 acceptance window. For 
the \PKzS\ and \PagL\ spectra, on the other hand, %are not compatible with flat distributions, but
we observe a rapidity dependence.    
The rapidity distributions for these particles are well described by Gaussians 
centred at mid-rapidity.  
%\newline
%The results of the $\chi^2$\ best-fit with either a Gaussian (\PKzS\ and \PagL) or a constant 
%(\PgL, \PgXm, \PagXp\ and \PgOm+\PagOp) are given in table~\ref{tab:InvMSD}.  
%
\begin{table}[t]
\caption{%Results of the maximum likelihood fits with a constant 
         %to the $dN/dy$\ distributions of \PgL, \PgXm, \PagXp\ and \PgOm+\PagOp. 
         Absolute yields in one unit of rapidity calculated by means of maximum  
         likelihood fits.  
         %~\cite{NA57Enh,EneDepPaper}.  
         The first error is statistical, the second is systematic.  
         Values of the $\chi^2/ndf$\ for the measured $dN/dy$\ distributions 
         %of \PgL, \PgXm, \PagXp\ and \PgOm+\PagOp\   
         calculated assuming  $dN/dy$\ constant and equal to the absolute yields   
         are also quoted.   
%$\frac{dN}{dy}$, $\sigma$ (for \PKzS, \PagL\ and \PagXp) and
\label{tab:InvMSDb}}
\begin{center}
\begin{tabular}{|c|cc|}
\hline
%   & $\frac{dN}{dy} \left|_{|y-y_{cm}|<0.5}  \right.$ 
   & $\int^{y_{cm}+0.5}_{y_{cm}-0.5}{\frac{dN}{dy}\, dy} $ 
   & $\chi^2/ndf$ \\ \hline
%\PKzS   &$11.7\pm0.3\pm1.2$ & $29/6$\\  %assuming dN/dy flat 
%\PKzS(new)   &$12.3\pm0.4\pm1.2$ & \\  %assuming dN/dy gaussian and T=237 MeV
\PKzS   &$11.8\pm0.3\pm1.2$ & $29/6$\\  %assuming dN/dy gaussian and T=248 MeV
%\PagL   &$1.17\pm0.03\pm0.12$& $13/6$\\ %assuming dN/dy flat
\PagL   &$1.14\pm0.03\pm0.12$& $12/6$ \\ %assuming dN/dy gaussian
\PgL    &$7.84\pm0.21\pm0.78$   & $3.9/6$ \\
\PgXm   &$0.852\pm0.015\pm0.085$ & $3.7/7$ \\
\PagXp  &$0.209\pm0.007\pm0.021$ & $3.6/5$ \\
\PgOm+
\PagOp  &$0.172\pm0.013\pm0.026$ & $5.0/4$ \\
\hline
%%%%%%%%%%%%%%%%%%%%%%%%%%%%%%%%%%%%%%%%%%%%%%%%%%%%%%%
\end{tabular}
\end{center}
\end{table}

The total yields integrated over one unit of rapidity as reported 
in references~\cite{NA57Enh,EneDepPaper}  are also shown in 
figure~\ref{fig:yspectra_160} 
as shaded bands superimposed to the corresponding 
rapidity distributions. In these bands the dotted and full lines represent, respectively, the 
statistical  and the systematic errors. %on the yields. 
These yields are calculated with a maximum likelihood fit   
using the Gaussian parameterization of $dN/dy$\ for \PKzS\ and \PagL, 
and a flat $dN/dy$\ for all other particles.  
The numerical values of the yields are reported  in table~\ref{tab:InvMSDb} 
along with their errors~\footnote{In previous conference proceedings and in 
reference~\cite{EneDepPaper} the $\frac{{\rm d}^2N}{{\rm d}p_{\tt T}  {\rm d}y }$\ distributions of 
 \PKzS\ and \PagL\ were %estrapolated 
evaluated  
assuming  flat rapidity distributions:  
%the difference is less than the statistical error.   
the resulting yields are equal to $11.7\pm0.3$\ and $1.17\pm0.03$, 
respectively; %are compatible within less than 1 $\sigma_{stat}$\   
the difference is less than the statistical error.   
%with those obtained  %here  
%assuming Gaussian distributions.
}.  
%The quoted yields of \PKzS\ and \PagL\ agree within the statistical errors with  
%the integrals of the best-fit Gaussian over the interval $-0.5<y-y_{cm}<0.5$,    
%equal to 12.1 and 1.15 for \PKzS\ and \PagL, respectively.   

The %results of the 
Gaussian parameters which minimize the $\chi^2$\ of the \PKzS\ and \PagL\ distributions are  
given in table~\ref{tab:InvMSD}.   
%%%%%%%%%%%%%%%%%%%%%%%%%%%%%%%%%%%%%
\begin{table}[h]
%\caption{Results of best-fits with either a Gaussian (\PKzS\ and \PagL) or
%         a constant (\PgL, \PgXm, \PagXp\ and \PgOm+\PagOp) to the rapidity distributions.
\caption{Results of fits with a Gaussian
        to the rapidity distributions of \PKzS\ and \PagL.
\label{tab:InvMSD}}
\begin{center}
\begin{tabular}{|c|ccc|}
\hline
   & $\frac{dN}{dy} \left|_{y=y_{cm}} \right.$\ & $\sigma$\ & $\chi^2/ndf$ \\ \hline
\PKzS   &$13.2\pm0.4$    & $0.69\pm0.07$  & $2.5/5$ \\
\PagL   &$1.22\pm0.04$   & $0.83\pm0.22$  & $3.0/5$ \\
%\PgL    &$7.75\pm0.19$   &                & $3.9/6$ \\
%%\PagL   &$1.22\pm0.04$   & $0.83\pm0.22$  & $3.0/5$ \\
%\PgXm   &$0.851\pm0.014$ &                & $3.5/7$ \\
%\PagXp  &$0.212\pm0.007$ &                & $3.2/5$ \\
%\PgOm+
%\PagOp  &$0.157\pm0.008$ &                & $4.9/4$ \\
\hline
%%%%%%%%%%%%%%%%%%%%%%%%%%%%%%%%%%%%%%%%%%%%%%%%%%%%%%%
\end{tabular}
\end{center}
\end{table}
%The quoted yields of \PKzS\ and \PagL\ agree %within the statistical errors
%with the integrals of the best-fit Gaussian over the interval $-0.5<y-y_{cm}<0.5$,  
%equal to 12.1 and 1.15 for \PKzS\ and \PagL, respectively.  
%Similarly, the $\chi^2$\ %best-fits 
%fits  with a constant to the rapidity distributions of \PgL, \PgXm, \PagXp, \PgOm\ 
%and \PagOp\  yield results fully compatible 
%with those %of references~\cite{NA57Enh,EneDepPaper} 
%reported in table~\ref{tab:InvMSDb},  
%where the more reliable maximum likelihood fits were used; in fact the values of the 
%$\chi^2/ndf$\ calculated assuming $dN/dy$ constant and equal to the yields 
%given in table~\ref{tab:InvMSDb} are found to be distributed around one.  
%These values are also given in the table.   
%
%Similarly,  the quoted yields of \PKzS\ and \PagL\ agree %within the statistical errors  
%with the integrals of the best-fit Gaussian over the interval $-0.5<y-y_{cm}<0.5$,
%equal to 12.1 and 1.15 for \PKzS\ and \PagL, respectively.   
\subsection{Comparison with NA49 results}
The large acceptance NA49 experiment has  
measured the rapidity distributions of a number of particle species over a broad rapidity 
range ($\approx 3$\ units of rapidity centred around mid-rapidity) in central Pb-Pb  
collisions~\cite{BlumeSQM04}.   
The \PgL\ rapidity distribution measured by NA49 is flat over the experiment's 
            $y$-coverage
            and the \PagL\ distribution has been fitted to a Gaussian with a
            width of $0.95\pm0.05$~\cite{NA49Lambda},
            compatible with our findings. 
NA49 has fitted the rapidity distributions of negatively and positively
         charged kaons with a two-Gaussian parameterization~\cite{NA49Kaons}.
         Assuming the \PKzS\ to be an average of \PKm\ and \PKp, and taking the
         (\PKm+\PKp)/2 average of the NA49 fitted parameters one obtains
         in our rapidity
         range ($2.93<y<3.70$) $rms=0.219$\  to be compared with  % da modificare (ask Orlando)
         $rms=0.221$\ of our measured distribution.
For the multi-strange $\Xi$\ and $\Omega$ hyperons, our rapidity 
distributions are compatible within the error bars both with being flat and with 
the shapes of the NA49 fits 
%to their distributions, 
%performed with either a single Gaussian or the sum of two Gaussian~\cite{NA49Xi,NA49Omega}.  
\cite{NA49Xi,NA49Omega}.

In conclusion, there is a fairly good agreement between the two experiments 
as far as the shape of the rapidity distributions are concerned.
%Therefore, as far as the shape of the rapidity distributions are concerned, there is a fairly 
%good agreement between the two experiments. 
On the other hand a systematic discrepancy 
on the values of the yields is found relative to the NA49 results, which are  
%yields being 
lower for all particles by  about 25-30\%~\cite{QMElia}.   
The origin of such a discrepancy has not yet been understood   
%notwithstanding  
despite  
the joint efforts of the two collaborations.  
\subsection{Centrality dependence}
\begin{figure}[p]
\centering
\resizebox{1.00\textwidth}{!}{%
\includegraphics{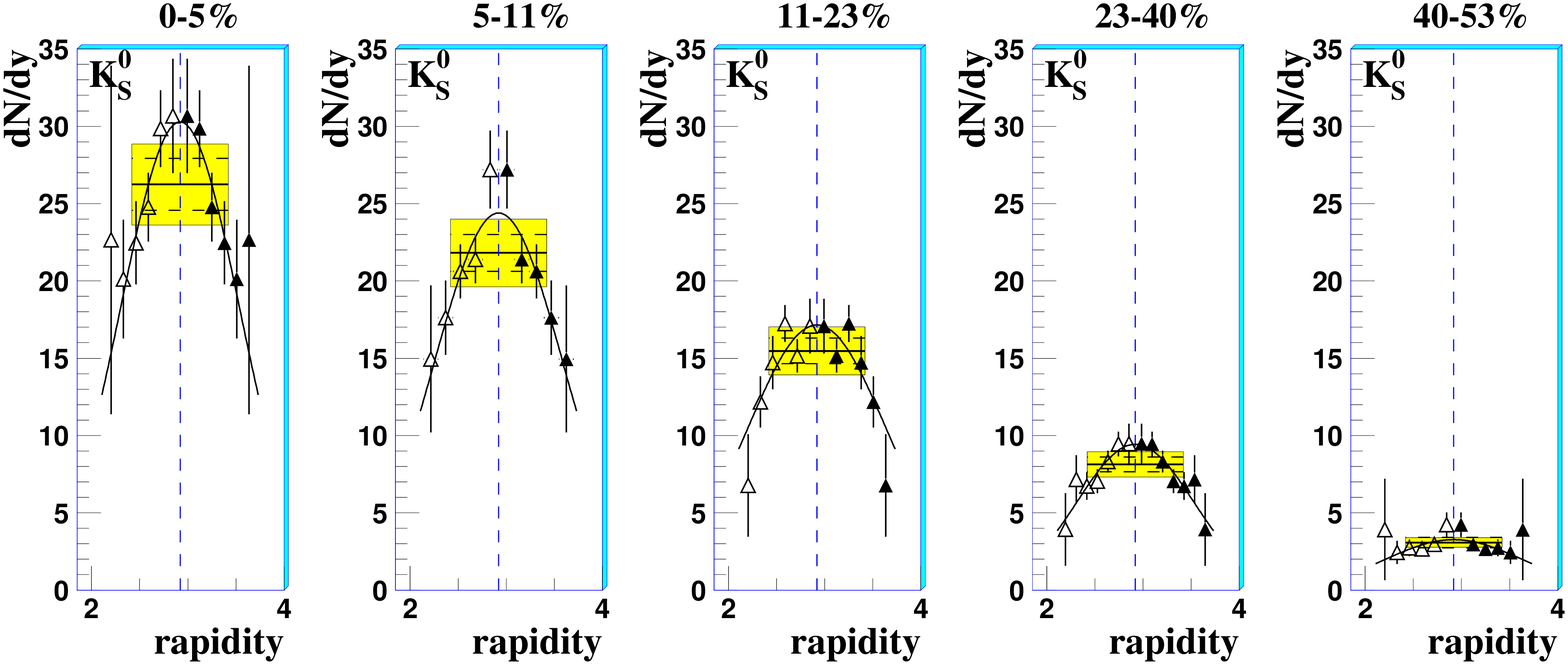}}
\caption{Rapidity distributions of \PKzS\ for the five centrality 
         %classes of table~\ref{tab:centrality}.
        classes.
\label{fig:K0s_msd}}
\vspace{0.7cm}
\resizebox{1.00\textwidth}{!}{%
\includegraphics{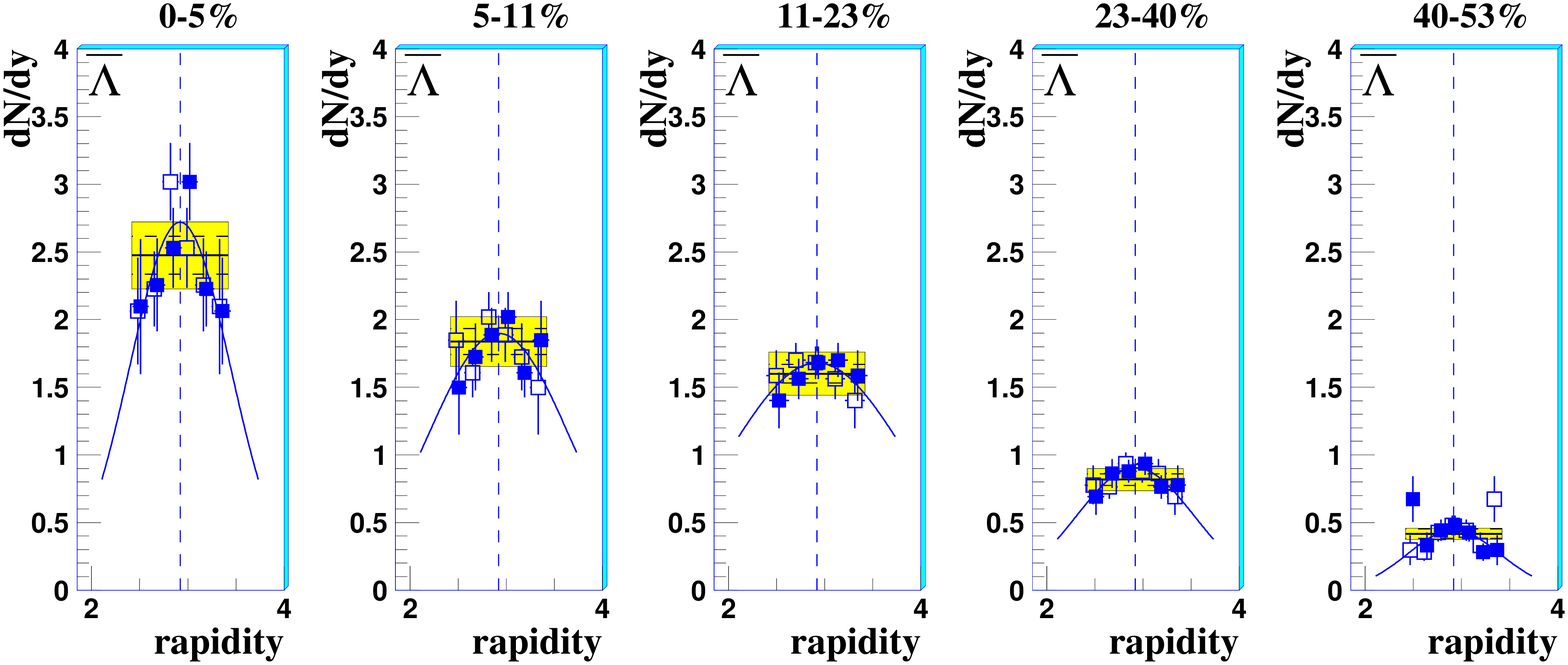}}
\caption{Rapidity distributions of \PagL\ for the five centrality 
         %classes of table~\ref{tab:centrality}.
          classes.
\label{fig:Al_msd}}
\vspace{0.7cm}
\resizebox{1.00\textwidth}{!}{%
\includegraphics{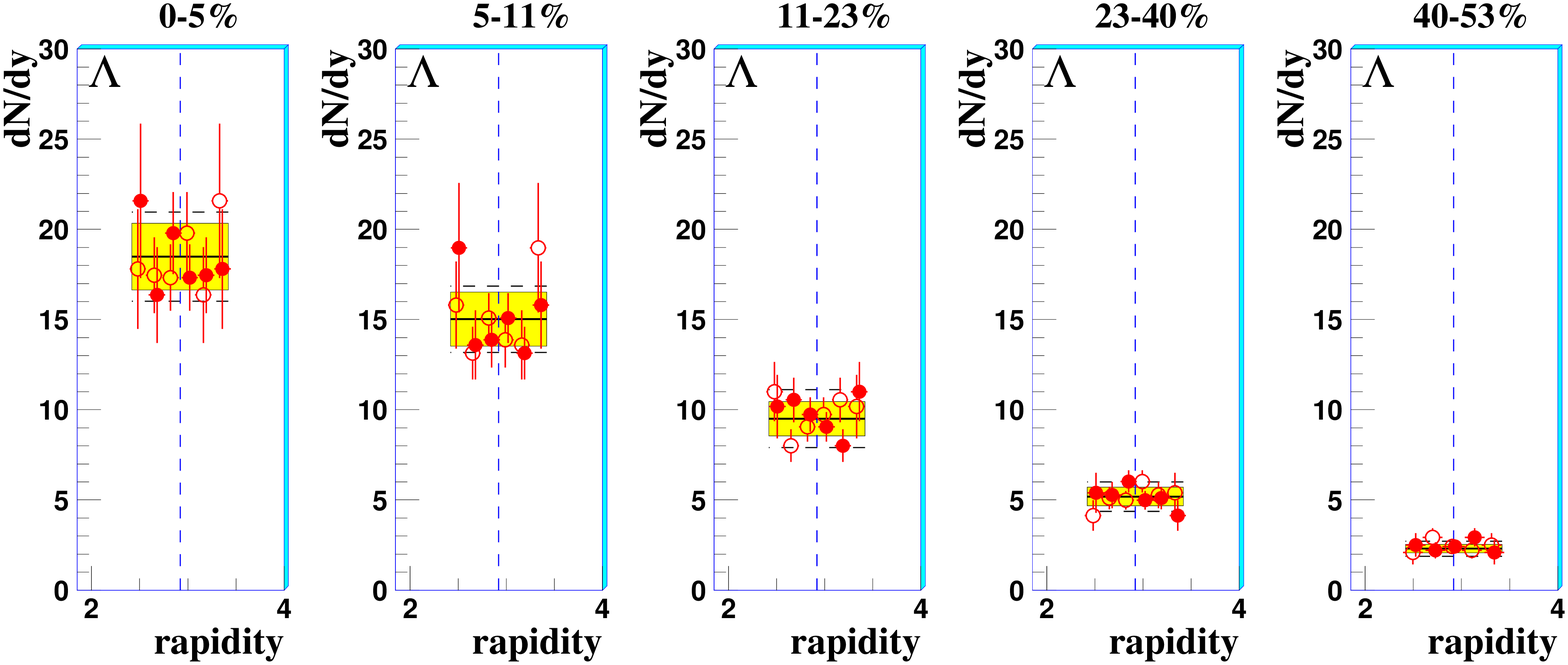}}
\caption{Rapidity distributions of \PgL\ for the five centrality
         %classes of table~\ref{tab:centrality}.
          classes.
\label{fig:La_msd}}
\end{figure}

In figures~\ref{fig:K0s_msd}~and~\ref{fig:Al_msd} we show the centrality dependence of the 
rapidity distributions for \PKzS\ and \PagL, respectively. As in figure~\ref{fig:yspectra_160}, 
the shaded bands represent our determination of the yield in one unit of rapidity, 
according to the maximum likelihood fit~\cite{NA57Enh,EneDepPaper}, 
with the dotted and full lines indicating 
the statistical and systematic errors, respectively.  
Gaussian $\chi^2$\ fits to the spectra are also superimposed,  with the fit parameters 
reported in table~\ref{tab:FitMSD}.   
For both particles, the width of the rapidity distributions is constant within the 
errors in the five centrality classes    
(i.e. from 40--53\% to 0--4.5\%, see table~\ref{tab:centrality}).  

\begin{figure}[p]
\centering
\resizebox{1.00\textwidth}{!}{%
\includegraphics{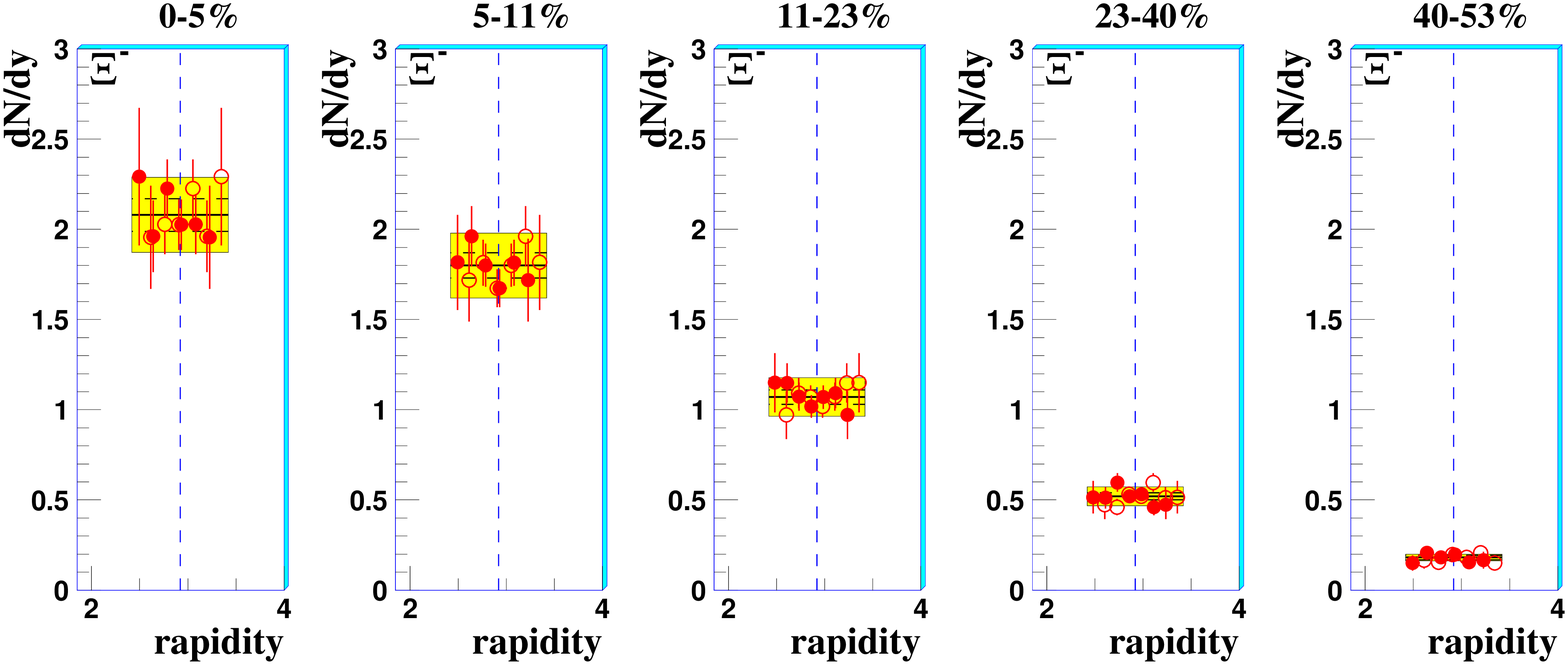}}
\caption{Rapidity distributions of \PgXm\ for the five centrality 
         %classes of table~\ref{tab:centrality}.
         classes.  
\label{fig:Xi_msd}}
\vspace{0.7cm}
\resizebox{1.00\textwidth}{!}{%
\includegraphics{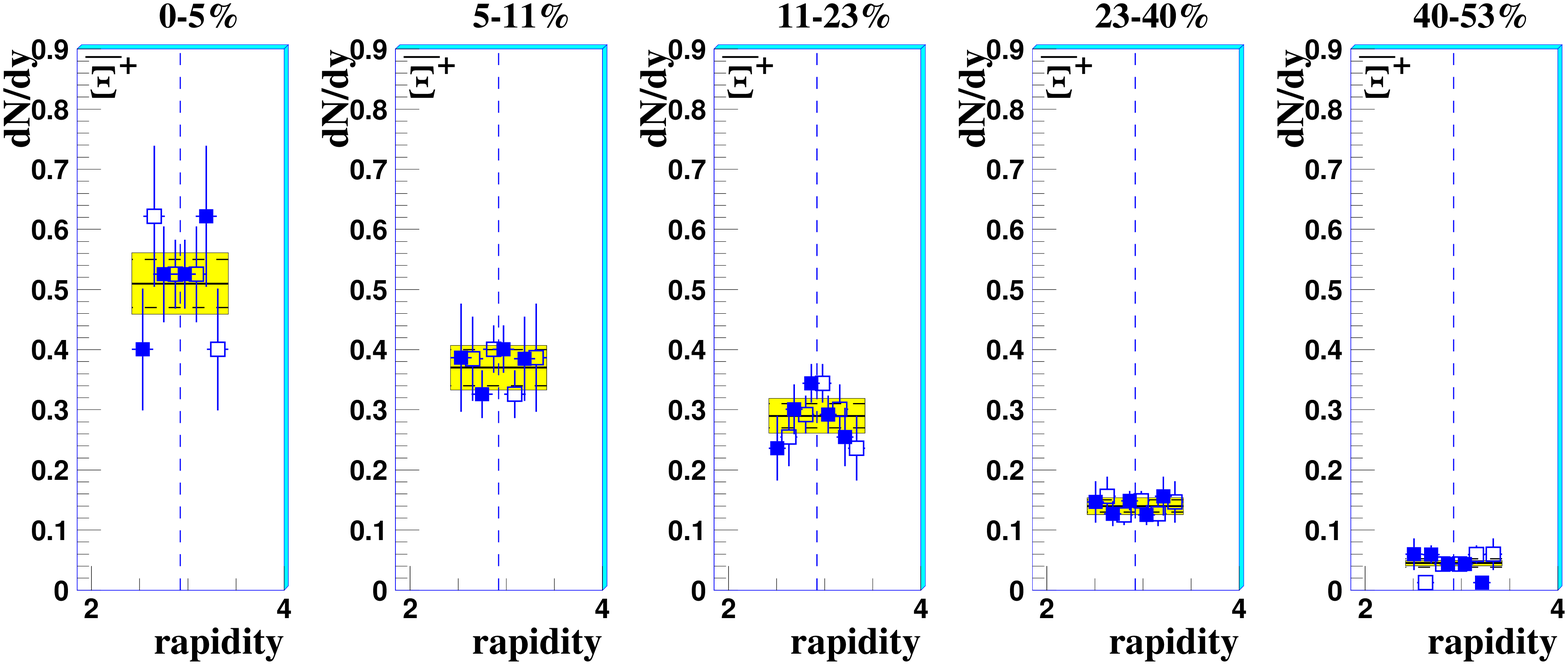}}
\caption{Rapidity distributions of \PagXp\ for the five centrality 
         %classes of table~\ref{tab:centrality}.
         classes.  
\label{fig:AXi_msd}}
\vspace{0.7cm}
\resizebox{1.00\textwidth}{!}{%
\includegraphics{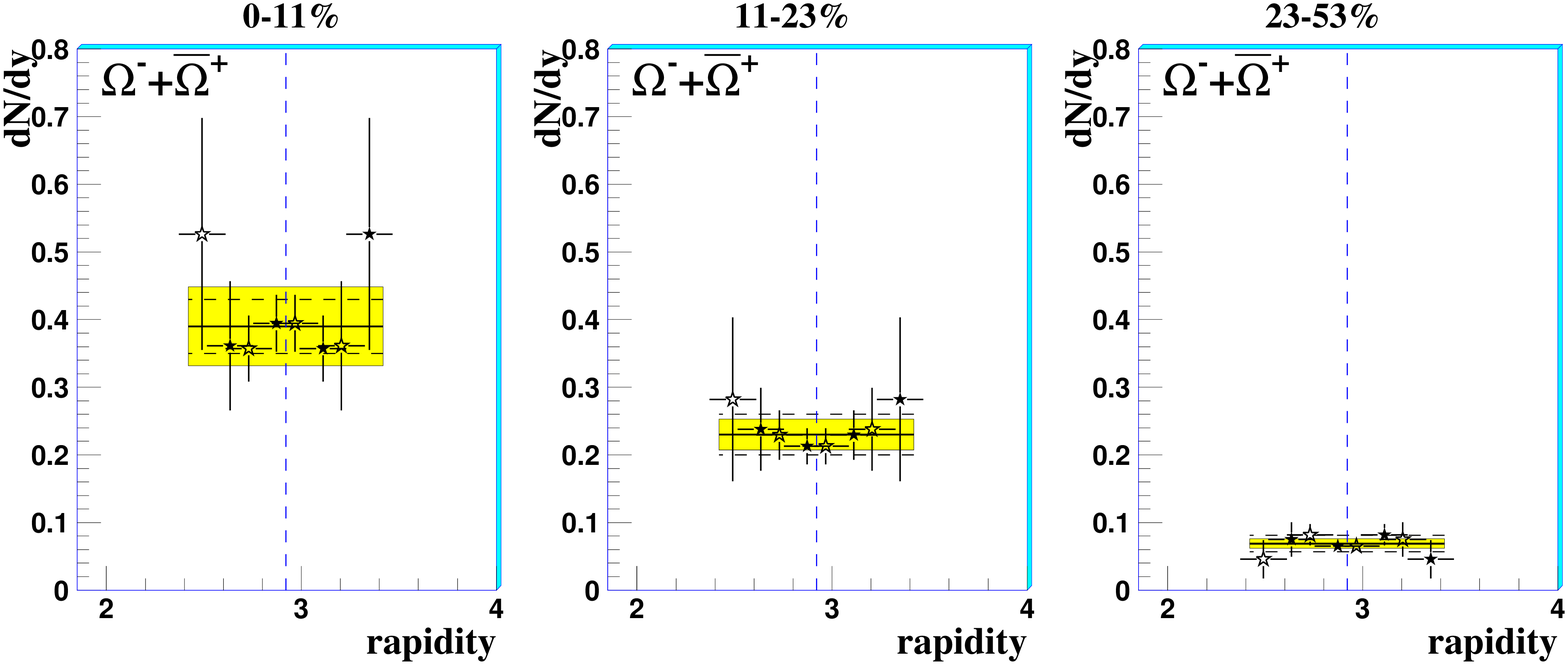}}
\caption{Rapidity distributions of \PgOm+\PagOp\ for central (0-11\%),
         semi-central (11-23\%) and semi-peripheral (23-53\%) collisions.  
\label{fig:OmAOm_msd}}
\end{figure}

The rapidity distributions in the five centrality classes  
of \PgL, \PgXm\ and \PagXp\ are shown in  
figures~\ref{fig:La_msd}, \ref{fig:Xi_msd} and \ref{fig:AXi_msd}, respectively. 
For statistical reasons
the rapidity distributions of \PgOm+\PagOp\ have been  
calculated in three centrality classes instead of five: $0+1$, $2$\ and $3+4$\ and
they are shown in figure~\ref{fig:OmAOm_msd}.  
In all the centrality classes, the rapidity distribution of the \PgL\ hyperon 
is consistent with being flat over the considered range.   
%, which cover about one unit of rapidity  about mid-rapidity.  
In the same rapidity range, the \PagL\ distribution varies by    
about 40\% (class 4).   
It is likely that the \PgL\ hyperon rapidity distribution reflects the overall  
net baryon number distribution.   
%hence the \PgL\ rapidity distribution would reflect the overall 
%net baryon number distribution,   
%which %is not peaked 
%does not show a peak  
%at mid-rapidity due to incomplete stopping  
%of the incoming nucleons at SPS energy. 
The same behaviour was observed for the $y$\  
distribution of protons in central Pb-Pb collisions at the same 
energy by the NA49 experiment~\cite{NA49netprot}.  
The \PgXm\ distributions are found to be flat in one   
unit of rapidity like the \PgL\ distributions.   
%thus suggesting that also the \PgXm\ hyperon %to have 
%conserves  
%``memory''  of the initial  baryon density.  
For the rare \PagXp and $\Omega$\ particles,  
the NA57 collected statistics do not allow us to 
observe significant deviations from a flat distribution in our limited rapidity range.  
\begin{table}[h]
\caption{Results of fits with a Gaussian to the rapidity distributions of \PKzS\ and \PagL\ 
         for the five centrality classes defined in table~\ref{tab:centrality}.
\label{tab:FitMSD}}
\begin{center}
\begin{tabular}{|ll|ccccc|}
\cline{3-7} \multicolumn{2}{c|}{ }
                     &   4      &    3      &    2      &    1    &    0   \\ \hline
\PKzS &$\frac{dN}{dy} \left|_{y=y_{cm}} \right.$
               &  $30.3\pm2.2$ & $24.4\pm1.6$ & $17.2\pm0.9$ & $ 9.4\pm0.6$ & $3.3\pm0.4$ \\
      &$\sigma$&  $0.61\pm0.12$& $0.67\pm0.13$& $0.72\pm0.13$& $0.61\pm0.11$& $0.71\pm0.34$\\
      &$\chi^2/ndf$&   $1.2/4$ &     $2.4/3$  &    $5.2/4$   &   $ 1.9/5 $  &   $2.4/4$ \\ \hline
\PagL &$\frac{dN}{dy} \left|_{y=y_{cm}} \right.$
               &  $2.72\pm0.20$ & $1.90\pm0.13$ & $1.68\pm0.09$ & $ 0.90\pm0.06$ & $0.44\pm0.05$ \\
      &$\sigma$&  $0.52\pm0.14$& $0.73\pm0.34$& $0.9\pm0.5$  & $0.61\pm0.20$& $0.48\pm0.19$\\
      &$\chi^2/ndf$&   $2.6/4$ &     $2.4/4$  &    $1.14/3$    &   $ 1.18/4 $  &   $7.0/5$ \\ \hline
\end{tabular}
\end{center}
\end{table}

The values of the yield in one unit of rapidity around mid-rapidity
as a function of the centrality
of the collision are given
in table~\ref{tab:FitMSDb} for all the particles.
\begin{table}[h]
\caption{Strange particle yields in one unit of rapidity around mid-rapidity,  
         % $\frac{dN}{dy} \left|_{|y-y_{cm}|<0.5} \right.$,  
         $\int^{y_{cm}+0.5}_{y_{cm}-0.5}{\frac{dN}{dy}\, dy} $,  
         for the five centrality classes defined in table~\ref{tab:centrality}. 
         Statistical (first) and systematic (second) errors are also quoted.   
         \vspace{0.3cm}
         %$\chi^2/ndf$\ 
\label{tab:FitMSDb}}
\footnotesize{
\hspace{-0.6cm}
\begin{tabular}{|l|cc|c|cc|}
\cline{2-6} \multicolumn{1}{c|}{ }
                     &   4      &    3      &    2      &    1    &    0   \\ \hline
% Assuming dN/dy flat
%\PKzS &$26.2\pm1.7\pm2.6$&$21.8\pm1.2\pm2.2$&$15.5\pm0.8\pm1.6$&$8.1\pm0.5\pm0.8$
%      &$3.08\pm0.33\pm0.31$ \\ \hline
% Assuming dN/dy gaussian with T=237 MeV, free sigma 
%\PKzS(new) &$27.4\pm1.9\pm2.7$&$23.1\pm1.4\pm2.3$&$16.0\pm0.9\pm1.6$&$8.5\pm0.6\pm0.8$
%      &$3.15\pm0.39\pm0.31$ \\ \hline
% Assuming dN/dy gaussian with T=237 MeV, fixed sigma (0.7)
%\PKzS(new) &$27.1\pm1.7\pm2.7$&$22.5\pm1.3\pm2.2$&$16.0\pm0.8\pm1.6$&$8.4\pm0.5\pm0.8$
%      &$3.18\pm0.34\pm0.31$ \\ \hline
% Assuming dN/dy gaussian with T=248.55 MeV, fixed sigma (0.7)
\PKzS &$26.0\pm1.7\pm2.6$&$21.6\pm1.2\pm2.2$&$15.3\pm0.8\pm1.6$&$8.1\pm0.5\pm0.8$
      &$3.05\pm0.33\pm0.31$ \\ \hline
% Assuming dN/dy flat
%\PagL &$2.47\pm0.14\pm2.5$&$1.84\pm0.10\pm0.18$&$1.60\pm0.07\pm0.16$&$0.82\pm0.04\pm0.08$
%      &$0.417\pm0.035\pm0.042$ \\ \hline
%Assuming dN/dy gaussian with T=287, free sigma
%\PagL(new) &$2.38\pm0.17\pm2.4$&$1.80\pm0.12\pm0.18$&$1.58\pm0.07\pm0.16$&$0.80\pm0.06\pm0.08$
%      &$0.41\pm0.04\pm0.04$ \\ \hline
%Assuming dN/dy gaussian with T=287, fixed sigma (0.8)
\PagL &$2.44\pm0.14\pm2.4$&$1.81\pm0.10\pm0.18$&$1.58\pm0.07\pm0.16$&$0.80\pm0.04\pm0.08$
      &$0.41\pm0.03\pm0.04$ \\ \hline
\PgL  
      & $18.5\pm1.1\pm1.8$ & $15.0\pm0.8\pm1.5$ & $9.5\pm0.5\pm0.9$ & $5.19\pm0.29\pm0.51$ 
      & $2.30\pm0.22\pm0.23$ \\
%      & $1.9/5$ & $3.2/5$   &    $4.7/5$  &    $3.7/5$  &  $1.5/4$ \\   %to be updated
\hline
\PgXm 
       &  $2.08\pm0.09\pm0.21$ & $1.80\pm0.07\pm0.18$ & $1.07\pm0.04\pm0.11$ & $0.52\pm0.02\pm0.05$ 
       & $0.181\pm0.013\pm0.018$ \\
%       & $1.9/5$ & $2.4/5$   &    $2.0/6$  &    $3.7/5$  &  $2.8/5$ \\   %to be updated
\hline
\PagXp  
       &$0.51\pm0.04\pm0.05$&$0.37\pm0.03\pm0.04$&$0.29\pm0.02\pm0.03$&$0.14\pm0.01\pm0.01$& 
       $0.045\pm0.007\pm0.004$\\
%       & $2.2/3$ & $1.9/3$ & $4.3/4$ & $1.7/4$ & $11.2/4$ \\  %to be updated
\hline
$\Omega$\   % now systematic error is 15%
 &  \multicolumn{2}{c|}{$0.39\pm0.04\pm0.06$} & $0.23\pm0.03\pm0.03$ &
       \multicolumn{2}{|c|}{$0.069\pm0.012\pm0.010$} \\
%& \multicolumn{2}{c|}{$1.0/3$} & $0.4/3$ & \multicolumn{2}{|c|}{$1.9/3$} \\ %to be updated
\hline
%%%%%%%%%%%%%%%%%%%%%%%%%%%%%%%%%%%%%%%%%%%%%%%%%%%%%%%
\end{tabular}
}
\end{table}
%
%A similar conclusion to that of the \PagXp %holds also 
%can be drawn 
%for the $\Omega$\ particle.  
%Due to the more limited statistics, 

%As for the yields in the integrated centrality range of table~\ref{tab:InvMSDb} these 
%yields 
%\section{Longitudinal flow}
\section{Longitudinal dynamics}

The {\it transverse} dynamics of the collisions have been studied in  
reference~\cite{BlastPaper} from the analysis of the transverse  
momentum distributions of the strange particles 
in the frame-work of the blast-wave model~\cite{BlastModel}.  
The rapidity distributions can be used to extract information about the 
{\it longitudinal} dynamics. We use 
an approach outlined in reference~\cite{BlastModel} 
(i.e., the same blast-wave model used for the study of the  
transverse dynamics) and~\cite{Indiani}, 
where, respectively, Bjorken and Landau  
hydrodynamics~\cite{Bjorken,Landau} are folded with a thermal  
distribution of the fluid elements.  

\begin{figure}[hbt]
\centering
\resizebox{0.80\textwidth}{!}{%
\includegraphics{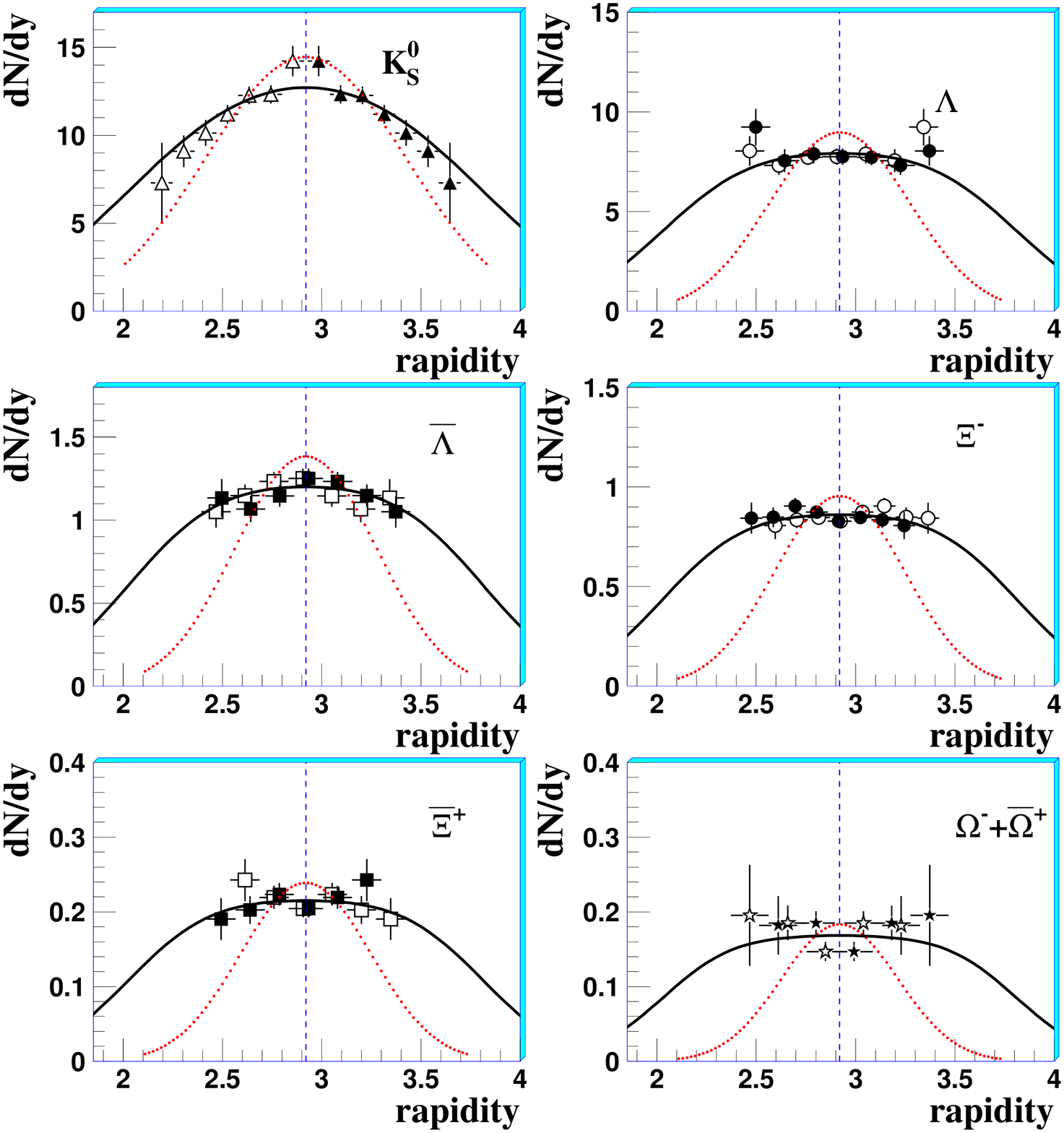}
}
\caption{Rapidity distributions of strange particles for the centrality range corresponding to 
          the most central 53\% of the inelastic Pb-Pb cross-section as compared to the thermal 
          model calculation of equation~\ref{eq:Bjorken1} (dotted lines, in red) and a thermal 
          model with longitudinal flow (full lines, in black).  
       }
\label{fig:Bjorken}
\end{figure}

In figure~\ref{fig:Bjorken}  
the observed rapidity distributions are compared with the expectation for a stationary thermal  
source %(thin lines) 
and with a longitudinally boost-invariant superposition of multiple  
%boosted individual 
isotropic, locally-thermalized sources %in a given rapidity interval 
(i.e. Bjorken hydrodynamics).    
Each locally thermalized source is modelled by an $m_{\tt T}$-integrated Maxwell-Boltzmann 
distribution with the rapidity dependence of the energy, $E=m_{\tt T} \cosh(\eta) $\ explicitly 
included  
\begin{equation}
\frac{dN_{th}}{d\eta}(\eta) = A T_f^3(\frac{m^2}{T_f^2}+\frac{m}{T_f}\frac{2}{\cosh \eta}
                      +\frac{2}{\cosh^2 \eta}) \exp(-\frac{m}{T_f}\cosh \eta)  \; ,
\label{eq:Bjorken1}
\end{equation}
where  $T_f$\ is the freeze-out temperature and $m_{\tt T}=\sqrt{p_{\tt T}^2+m^2}$.
The distributions are integrated over source element rapidity to extract the maximum 
longitudinal flow, $\eta_{max}$
\begin{equation}
 \frac{dN}{dy}=\int^{\eta_{max}}_{-\eta_{max}}{\frac{dN_{th}}{d\eta}(y-\eta) \,d\eta} \newline
 \beta_{L} = \tanh{\eta_{max}}
 \label{eq:Bjorken2}
\end{equation}
where $\beta_{L}$\ is the maximum longitudinal velocity in units of $c$. The average 
longitudinal flow velocity is evaluated as $<\beta_{L}> = \tanh(\eta_{max}/2)$.  
A simultaneous fit of the function defined by equation~\ref{eq:Bjorken2} to the rapidity distributions 
of the strange particles gives $<\beta_L> = 0.42\pm0.03$\ with $\chi^2/ndf=28.2/32$. The freeze-out  
temperature has been fixed to the value $T_f=144$\ MeV obtained, for the most central 53\% of 
the inelastic Pb-Pb collisions, from the analysis of the transverse mass spectra of the same 
group of particles~\cite{BlastPaper}.  In the same analysis the 
average {\em transverse} flow velocity has been determined to be $<\beta_{\perp}>=0.38\pm0.02$, i.e.  
slightly less than the {\em longitudinal} velocity determined in this analysis; this 
%finding would imply a quite large 
indicates substantial 
stopping of the incoming nuclei as a consequence of the collision.  

In principle, also the freeze-out temperature can be fitted from the rapidity distributions 
along with the longitudinal velocity. However, the sensitivity to the freeze-out temperature is very 
limited. For instance, changing $T_f$\ from 144 to 120 MeV results in only a 2\% increase 
of $<\beta_L>$.  
%but with poor sensitivity: e.g., with $T_f$\ fixed to 
%120\ MeV, one would get $<\beta_L>=0.41\pm0.02$.   
The partial contributions %in the global fit 
to the total $\chi^2$\ from the individual particle  
rapidity spectra are given in table~\ref{tab:Bjorken}.  
Within our uncertanties, we do not observe any particle to deviate 
from the common description given by a collective longitudinal flow 
superimposed to the thermal motion. A combined fit performed only to the \PKzS\ and \PagL\ 
rapidity distributions yields a smaller value of the flow, i.e.  $<\beta_{L}> \simeq 0.36\pm0.03$.  
It is worth noting that the flattening of the rapidity spectra with 
increasing  
particle mass, which is also observed in the data, is due in the model to the collective 
dynamics: %in this description: 
all particles are driven by the flow with the same velocity independently of 
their masses.  
\begin{table}[h]
\caption{Partial contributions to the $\chi^2$\ of the best-fit to the thermal 
         model with longitudinal flow, equation~\ref{eq:Bjorken2}. 
         The $\chi^2/ndf$\ of the global fit is 28.2/33, since there are six normalization 
         constants (the factors $A$\ in equation~\ref{eq:Bjorken2}) and the parameter   
         $\beta_L$.   
\label{tab:Bjorken}}
\begin{center}
\begin{tabular}{lll}
\hline
 particle & n. of points   & $\chi^2$    \\ \hline
 \PKzS    &  7 & 5.1  \\
 \PagL    &  7 & 3.6  \\
 \PgL     &  7 & 6.7  \\
 \PgXm    &  8 & 4.5  \\
 \PagXp   &  6 & 3.1  \\
 $\Omega$ &  5 & 5.2 \\ \hline 
 tot      & 40 & 28.2 \\ 
 \hline
\end{tabular}
\end{center}
\end{table}

In Landau hydrodynamics, the amount of entropy ($dS$) contained within a (fluid) rapidity $d\eta$\ 
is given by~\cite{Landau}
\begin{equation}
\frac{dS}{d\eta} 
%\propto -
= \pi R^2 l s_0 \; 
 \beta c_s \exp[\beta \omega_f]
  \left[ I_0(q) - \frac{\beta\omega_f}{q}I_1(q)  \right]
\label{eq:Landau}
\end{equation}
where $q=\sqrt{\omega_f^2-c_s^2\eta^2}$, $\omega_f=\ln(T_f/T0)$, $c_s$\ is the velocity of 
sound in the medium, %$T_f$\ is the freeze-out temperature, 
$T_0$\ is the initial temperature,
$\eta$\ is the rapidity, 
$R$\ is the radius of the nuclei, $2l$\ is the initial length, 
$s_0$\ is the initial entropy density, $2\beta=(1-c_s^2)/c_s^2$ and $I_0$, $I_1$\ 
are Bessel functions. The quantity $\pi R^2 l s_0$\ is used to normalize the 
spectra at mid-rapidity.  
The particle rapidity distribution is obtained, as for the Bjorken case, as a superpositon 
of the multiplicity density in rapidity space ($dN/d\eta \propto dS/d\eta$) %to the 
with a  
thermal distribution of the fluid elements, 
\begin{equation}
\frac{dN}{dy} \propto \int \frac{dN}{d\eta} \frac{dN_{th}}{d\eta}(y-\eta) \, d\eta
\label{eq:Landau2}
\end{equation}

\begin{figure}[hbt]
\centering
\resizebox{0.55\textwidth}{!}{%
\includegraphics{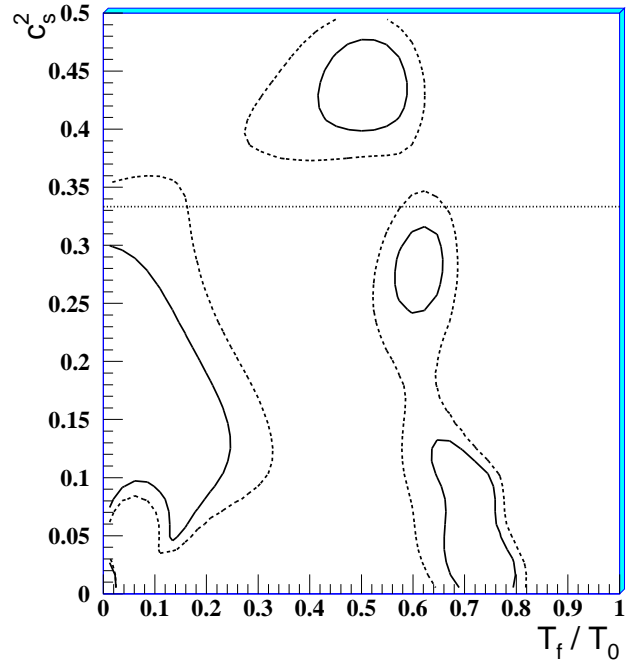}} 
\caption{The square of the velocity of sound in the medium %($c_s^2$, in unit of $c^2$) 
         (in unit of $c^2$)
         versus the ratio of the freeze-out temperature to the inital temperature. % ($T_f/T0$). 
         The 1$\sigma$\ (full curves) and the 3$\sigma$\ (dashed curves) confidence 
         contours are shown. The dotted line at $c_s^2=1/3$\ show the ideal gas limit.
       }
\label{fig:Landau}
\end{figure}

In the Landau model the width of the rapidity distribution is 
sensitive to the velocity of sound and to the ratio of the freeze-out temparature 
to the initial temperature. While integrating over $\eta$\ in equation~\ref{eq:Bjorken2}, 
the range of $\eta$\ is treated  as a parameter in case of Bjorken hydrodynamics; 
moreover in the Bjorken case (equation~\ref{eq:Bjorken2})  
the factor $dN/d\eta$, which  appears  in equation~\ref{eq:Landau2},   
%of Bjorken (equation~\ref{eq:Bjorken2}) 
has been included in the overall normalization factor $A$\ 
since the entropy density $dS/d\eta$\ is independent of the rapidity, in accordance with the 
assumption of boost invariance along the longitudinal direction~\cite{Bjorken}.   
In the case of Landau hydrodynamics, the integration limit is fixed by  
$\eta_{max}=-\eta_{min} =\ln(T_0/Tf)/c_s$~  
%$T_0/T_f=\exp(c_s \eta_{max})$  
\footnote{In reference~\cite{Srivastava} 
a modification has been developed ({\em Srivastava})  where the integration limit 
for rapidity is infinite, but this case has not been considered in 
the present analysis.} and the multiplicity density in $\eta$\ space ($dN/d\eta$) 
is written explicitly in the $\eta$\ integration (equation ~\ref{eq:Landau2}).   
Landau hydrodynamics can also reproduce simultaneously the distributions for all 
the  strange particles considered ($\chi^2/ndf \simeq$ 28/32) but we are not able 
to put stringent %and unique  
constraints on both the velocity of sound and the ratio $T_f/T_0$.  
The confidence level contours in the $c_s^2$\ vs. $\frac{T_f}{T_0}$\  
parameter space are shown in figure~\ref{fig:Landau}.  
For instance, the hypothesys of a perfect gas (i.e. $c_s^2=1/3$) would result  
(at the 3$\sigma$\ confidence level) in either $T_f/T_0 \approx 0$\ or  
$T_f/T_0 \approx 0.6$. In fact, two physical regions are constrained at 
the 3$\sigma$\ confidence level, the first located at small values of $T_f/T_0$\ and 
the second between 0.5 and 0.8; on the other hand, the region at $c_s^2>1/3$\ is 
unphysical.     
%unlikely; 
Both physical regions span over the full range $0<c_s^2<1/3$.   

\section{Conclusions}
We have measured the $dN/dy$\ distributions of %high statistics, 
high purity samples of \PKzS, \PgL, $\Xi$\ and $\Omega$\ particles produced 
at central rapidity in Pb-Pb collisions at 158 $A$\ GeV/$c$\ over a wide centrality 
range of collision (i.e. the most central 53\% of the Pb-Pb  
inelastic cross-section). 
%These data are well suited to be fitted to thermal models.  

%In spite of the narrow NA57 rapidity coverage 
%(about one unit of rapidity around mid-rapidity), 
In the unit of rapidity around mid-rapidity covered by NA57, 
we have performed fits to  
the $dN/dy$\ distributions of \PKzS\ and \PagL\ using  
a Gaussian parameterization: the resulting widths are compatible with each other 
%within the errors 
and constant as a function of centrality.   

Contrary to \PagL, the \PgL\ spectra are flat to good accurancy in the  
range of rapidity and centrality considered; this would indicate that the   
\PgL\ hyperon conserves ``memory'' of the initial baryon density.  
%carries a sizeable fraction of the total net baryon number.  
%reflects 
%From NA57 data a similar conclusion cannot be excluded for the \PgXm\ hyperon. %from our data.  
%The data suggest a similar interpretation for the \PgXm\ hyperon.   

The rapidity distributions of the $\Omega$\ particle are found to be 
flat within the errors  in one unit of rapidity for central (0-11\%)   
and peripheral (23-53\%) collisions.  

Boost-invariant Bjorken hydrodynamics can describe 
simultaneously the rapidity spectra of all the strange 
particles under study with $\chi^2/ndf \approx 1$, yielding an average longitudinal flow velocity  
$<\beta_L>=0.42\pm 0.03 $, sligthly larger than the measured transverse flow.  The {\em isotropic}   
collective expansion of the system suggests large  
nuclear stopping. % at SPS energy.  

A fairly good description is also provided by Landau hydrodynamics, 
%description, 
which allows us to put constraints in the parameter space of 
the velocity of sound in the medium and the ratio of the 
freeze-out temperature to the initial temperature.  
%
%%%%%%%%%%%%%%%%%%%%%%%%%%%%%%%%%%%%%%%%%%%%%%%%%%%%
%
%%%%%%%%%%%%%%%%%%%%%%%%%%%%%%%%%%%%%%%%%%%%%%%%%%%%%%%%%%
\ack
We are grateful to Jan-e Alam  
for useful comments and fruitful discussions.   
%%%%%%%%%%%%%%%%%%%%%%%%%%%%%%%%%%%%%%%%%%%%%%%%%%%%%%%%%%%%%%%%%%%%%%%%%
%%%%%%%%%%%%%%%%%%%%%%%%  BIBLIOGRAFIA %%%%%%%%%%%%%%%%%%%%%%%%%%%%%%%%%%

%%%

\section*{References}
\begin{thebibliography}{33}
%
%%%%%%%%%%%%%%%%%%%%%%%%%% Intro %%%%%%%%%%%%%%%%%%%%%%%%%%
%		   see also www.cern.ch/CERN/Announcements/2000/NewStateMatter
\bibitem{lattice} Karsch F 2002 {\it Lect. Notes Phys.} {\bf 583} 209
\bibitem{QM04} Quark Matter Conference Proceedings 2004 \JPG {\bf 30} S633-S1430
%\bibitem{QGP} Collins J C and Perry M J 1975 \PRL {\bf 34} 1352 \nonum
%              Cabibbo N and Parisi G 1975 \PL B {\bf 59} 67 \nonum
%              Karsch F 2002 \NP A {\bf 698} 199c 
%
\bibitem{NA57proposal} Caliandro R {\it et al.}, NA57 proposal, 1996
{\it CERN/SPSLC 96-40, SPSLC/P300}
\bibitem{StrEnh} Rafelski J and M\"{u}ller B 1982 \PRL {\bf 48} 1066  \nonum
                 Rafelski J and M\"{u}ller B 1986 \PRL {\bf 56} 2334
\bibitem{WA97Enh} Andersen E {\it et al.} 1999 \PL B {\bf 449} 401  \nonum
                  Antinori F {\it et al.} 1999 \NP A {\bf 661} 130c
\bibitem{NA57Enh} Bruno G E {\it et al.} 2004 \JPG {\bf 30} S717-S724 
%\nonum             Antinori F {\it et al.} 2005 {\it submitted to} \JPG
%
\bibitem{BlastPaper} Antinori F {\it et al.} 2004 \JPG {\bf 30} 823-840 
\bibitem{RcpPaper} Antinori F  {\it et al.} 2005 \PL B {\bf 623} 17-25
\bibitem{Busza} Busza W and Goldhaber A 1984 \PL B {\bf 139} 235
\bibitem{NA49netprot} Appelsh\"{a}user H {\it et al.} 1999 \PRL {\bf 82} 2471
%
\bibitem{Landau} Landau L D 1953  {\it Izv. Akad. Nauk.} SSSR {\bf 17} 51 \nonum
                 Belenkij S and Landau L D 1955 {\it Usp. Fiz. Nauk.} {\bf 56} 309 \nonum
                 Belenkij S and Landau L D 1956 \NC (suppl.) {\bf 3} 15
\bibitem{Bjorken} Bjorken J D 1983 \PR D {\bf 27} 140
%
%\bibitem{Horn} Ga\'{z}dzicki M  {\it et al.} 2004 \JPG {\bf 30} S701
%\bibitem{ThermalModel1} Braun-Munzinger P, Magestro D, Redlich K and Stachel J 
%                        2001 \PL B {\bf 518} 41
%\bibitem{ThermalModel2} Florkowsli W, Broniowski W and Michalec M 2002 
%                        {\it Acta Phys. Pol.} B {\bf 33} 761
%\bibitem{ThermalModel3} Rafelski J and Letessier J 2003 \NP A {\bf 715} 98
%\bibitem{ThermalModel4} Beccatini F, Gazdzicki M, Manninen J and Stock R 2004 
%                        \PR C {\bf 69} 024905
%\bibitem{Antinori} Antinori F 2004 \JPG {\bf 30} S725-S734
\bibitem{NA57descr} Manzari V {\it et al.} 1999 \JPG {\bf 25} 473 \nonum
		    Manzari V {\it et al.} 1999 \NP A {\bf 661} 761c
% Data sample
\bibitem{Multiplicity} Carrer~N {\it et al.} 2001 \JPG {\bf 27} 391 \nonum
 Antinori F {\it et al.} 2005 \JPG {\bf 31} 321-335
% Analysis and results 
\bibitem{GEANT} GEANT, CERN Program Library Long Writeup W5013
\bibitem{Fanebust} Fanebust K {\it et al.} 2002 \JPG {\bf 28} 160
\bibitem{BlastModel} Schnedermann E, Sollfrank J and Heinz U 1993 \PR C {\bf 48}  2462-2475
\bibitem{EneDepPaper} Antinori F  {\it et al.} 2004 \PL B {\bf 595} 68-74
%
\bibitem{BlumeSQM04} Blume C  {\it et al.} 2005 \JPG {\bf 31} S685-S691
\bibitem{NA49Lambda} Anticic T  {\it et al.} 2004 \PRL {\bf 93} 022302
\bibitem{NA49Kaons} Afanasiev S V {\it et al.} 2002 \PR C {\bf 66} 054902
\bibitem{NA49Xi} Afanasiev S V {\it et al.} 2002 \PL B {\bf 538} 275-281
\bibitem{NA49Omega} Alt C {\it et al.} Alt C {\it et al.} 2005 \PRL {\bf 94} 192301
%
\bibitem{QMElia} Elia D  {\it et al.} 2004 \JPG {\bf 30} S1329-S1332
%
%\bibitem{Protons} Appelsh\"{a}user H et al. 1999 \PRL {\bf 82} 2471
%
\bibitem{Indiani} Mohanty B and Alam J 2003 \PR C {\bf 68} 064903
%\bibitem{Bjorken} Bjorken J D 1983 \PR D {\bf 27} 140
%\bibitem{Landau} Landau L D 1953  {\it Izv. Akad. Nauk.} SSSR {\bf 17} 51 \nonum
%                 Belenkij S and Landau L D 1955 {\it Usp. Fiz. Nauk.} {\bf 56} 309 \nonum
%                 Belenkij S and Landau L D 1956 \NC (suppl.) {\bf 3} 15 
\bibitem{Srivastava} Srivastava D K, Alam J, Chakrabarty S, Raha S and Sinha B 1993 {\it Ann. Phys.} 
{\bf 228} 104
%
%%%%%%%%%%%%%%%% Rapidity distributions
%
%%%%%%%%%%%%%%
\end{thebibliography}
\end{document}